\documentclass[atoms,article,accept,pdftex,moreauthors]{Definitions/mdpi} 
\firstpage{1} 
\pubvolume{11}
\issuenum{6}
\articlenumber{89}
\pubyear{2023}
\copyrightyear{2023}
\externaleditor{Academic Editors:  J. Tito Mendonca and Hugo Terças}
\datereceived{19 April 2023} 
\daterevised{24 May 2023}
\dateaccepted{26 May 2023} 
\datepublished{31 May 2023} 
\hreflink{https://doi.org/\linebreak10.3390/atoms11060089} 

\usepackage[final]{changes}

\Title{Signatures of Quantum Chaos of Rydberg-Dressed Bosons in a Triple-Well Potential} 

\TitleCitation{{Signatures} of Quantum Chaos of Rydberg-Dressed Bosons in a Triple-Well Potential} 

\Author{{Tianyi} 
 Yan $^{1}$\orcidA{}, Matthew Collins $^{1}$, Rejish Nath {$^{2}$}\orcidB{} and Weibin Li {$^{1,3,}$}*\orcidC{}}

\AuthorNames{Tianyi Yan, Matthew Collins, Rejish Nath and Weibin Li}

\AuthorCitation{Yan, T.; Collins, M.; \mbox{Nath, R.}; Li, W.}

\address{%
$^{1}$ \quad School of Physics and Astronomy, University of Nottingham, Nottingham NG7 2RD, UK; tianyi.yan@nottingham.ac.uk (T.Y.); {ppymlc@exmail.nottingham.ac.uk} (M.C.)\\ 
$^{2}$ \quad Department of Physics, Indian Institute of Science Education and Research, {Dr. Homi Bhabha Road, } 
Pune~411008, {India}; rejish@iiserpune.ac.in\\ 
$^{3}$ \quad Centre for the Mathematics and Theoretical Physics of Quantum Non-Equilibrium Systems, \linebreak University of Nottingham, Nottingham NG7 2RD, UK

}

\corres{Correspondence: weibin.li@nottingham.ac.uk}

\abstract{We studied signatures of quantum chaos in dynamics of Rydberg-dressed bosonic atoms held in a one-dimensional triple-well potential. Long-range nearest-neighbor and next-nearest-neighbor interactions, induced by laser dressing atoms to strongly interacting Rydberg states, drastically affect mean-field and quantum many-body dynamics. By analyzing the mean-field dynamics, classical chaos regions with positive and large Lyapunov exponents were identified as a function of the potential well tilting and dressed interactions. In the quantum regime, it was found that level statistics of the eigen-energies gain a Wigner--Dyson distribution when the Lyapunov exponents are large, giving rise to signatures of strong quantum chaos. We found that both the time-averaged entanglement entropy and survival probability of the initial state have distinctively large values in the quantum chaos regime. We further showed that population variances could be used as an indicator of the emergence of quantum chaos. This might provide a way to directly probe quantum chaotic dynamics through analyzing population dynamics in individual potential wells. }

\keyword{{Rydberg dressing; Bose--Hubbard model; quantum chaos; level statistics; entanglement entropy}} 

\begin{document}




\section{Introduction}

{Understanding} 
 dynamics of quantum many-body systems has been a lucrative field of study that allows for the exploration of new physics and finding quantum technological applications. Among~many experimental platforms, ultracold atoms trapped in optical lattices, due to their high controllability, provide a versatile toolbox for studying quantum many-body phases~\cite{bloch_many-body_2008}. One is typically keen to achieve long-range interactions, which allow us to create and probe exotic many-body dynamics. A~paradigmatic example is the extended Bose--Hubbard model (EBHM), where rich phases (e.g., superfluid, supersolid and checkboard phases) are obtained~\cite{PhysRevLett.88.170406,Trefzger_2011,Rossini_2012, Ejima2014SpectralAE,PhysRevA.88.063608}. Dipolar atoms provide relatively weak long-range interactions~\cite{chomaz_dipolar_2022}.  
 Recent experiments have shown that much stronger long-range interactions can be achieved in electronically high-lying Rydberg states~\cite{saffman_quantum_2010}. Long coherence times are realized through Rydberg dressing, where ground-state atoms are weakly coupled to Rydberg states using far off-resonant lasers~\cite{Bouchoule2002}. Such Rydberg dressing leads to a soft-core-shaped long-range interaction potential between ground-state atoms, where the radius of the soft-core potential is in the order of a few micrometers~\cite{Bouchoule2002,PhysRevLett.104.195302,Honer2010,Pupillo2010,Johnson2010, PhysRevA.85.053615,Xiong2014,Hsueh2020}. 
 
\begin{figure}[H]

    \includegraphics[width=0.8\linewidth]{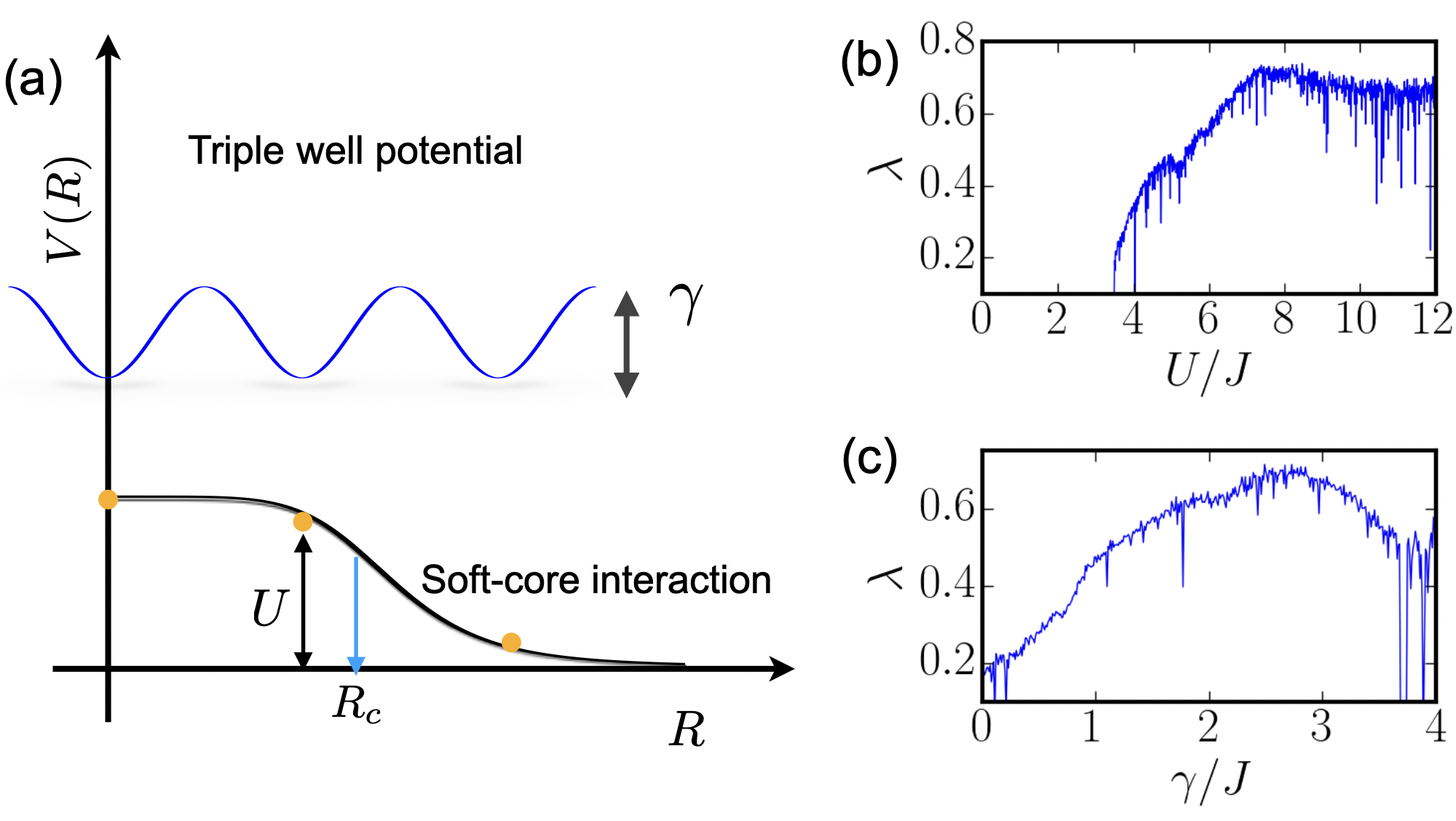}
    \caption{(\textbf{a}) {Three}-well trapping potential and soft-core-shaped interaction between Rydberg-dressed atoms. The~soft-core radius $R_c$ is larger than the distance between neighboring sites. By~tuning $R_C$, both the nearest-neighbor and next-nearest-neighbor  interaction can be made to be strong.  \mbox{(\textbf{b},\textbf{c})}~Lyapunov exponents in the semiclassical regime. Parameters are $\gamma/J = 2.5$ and $U/J=7.0$ in (\textbf{b}) and (\textbf{c}), respectively. See text for details.} 
    \label{potential}
\end{figure}

Existing studies have focused on static and dynamical properties of Rydberg-dressed atoms confined in traps~\cite{Maucher2011c,Cinti2014,mukherjee_phase-imprinting_2015,Hsueh2016,McCormack2020,li_many-body_2020,zhou_multipolar_2021} and optical lattices~\cite{Lauer2012,Lan2015,Angelone2016,Chougale2016,Li2018a,Zhou2020,PhysRevA.99.033602,McCormack2020b}. 
Interaction effects due to the Rydberg dressing have been experimentally demonstrated in optical tweezers~\cite{Jau2016b}, optical lattices~\cite{Zeiher2016,Zeiher2017,Guardado-Sanchez2020} and~harmonic traps~\cite{Borish2020}.  In~optical lattice potentials, such a large soft-core radius is typically much larger than the lattice spacing, as~depicted in Figure~\ref{potential}a. The dynamics of Rydberg-dressed bosons are described by an EBHM~\cite{PhysRevA.85.053615}.  In~Refs.~\cite{McCormack2020b,mccormack_hyperchaos_2021}, we studied nonlinear and chaotic dynamics of Rydberg-dressed bosons in finite  potential wells in the semiclassical regime.  
In the quantum regime, it is known that the BHM is non-integrable in general, where quantum chaos can be triggered by the competition between coherent hopping and on-site interactions~\cite{PhysRevE.102.032208,CHOY198283,Kolovsky_2004,PhysRevB.75.115119,PhysRevE.103.042109}. In~the presence of long-range interactions, it has been shown that quantum chaos emerges in EBHMs realized with dipolar bosons, which feature nearest-neighbor interactions~\cite{PhysRevE.105.034204,Kollath_2010}. The~respective chaotic properties are characterized by quantities such as level statistics~\cite{Kollath_2010}, Shannon entropy, and~survival probabilities of the initial state~\cite{PhysRevE.105.034204}.
\vspace{-6pt}


In this work, we investigated signatures of the quantum chaos of Rydberg-dressed bosons in finite one-dimensional traps. Because~of the large soft-core radius, our setup allowed us to focus on the dynamics in a region where both nearest-neighbor and next-nearest-neighbor interactions are equally strong. In~the semiclassical regime, Lyapunov exponents of the mean-field equations were evaluated numerically. Positive and large-valued Lyapunov exponents were found, signifying the emergence of chaos. The dependence of the Lyapunov exponents on the dressed interaction and tilting of the potential was studied. By~diagonalizing quantum many-body Hamiltonians with finite atom numbers, we show that nearest-neighbor level statistics gain a Wigner--Dyson distribution with parameters when the Lyapunov exponents are large. We further characterized signatures of quantum chaos through analyzing quantum many-body dynamics. It was found that the time-averaged entanglement entropy shows a similar dependence on the parameters to that of the level statistics and Lyapunov exponents. Features of quantum chaos can be captured by variances of the population in different potential wells, which might provide a way to directly probe the quantum~chaos.  

The structure of the paper is as follows. In~Section~\ref{model}, we introduce the system and model Hamiltonian. Section~\ref{result} is divided into three sub-sections. In~Section~\ref{levelstat}, the~level statistics of the system are studied. Parameters corresponding to Wigner--Dyson distribution are found. In~Section~\ref{eentropy}, dynamics of the entanglement entropy are investigated. The~time-averaged entanglement entropy gains larger values in the quantum chaos regime. In~Section~\ref{surProb}, we evaluate the survival probability of the initial state and particle numbers in different potential wells. Population variance in different regimes is examined. Conclusions are given in Section~\ref{conclusion}.

\section{Model}
\label{model}

We considered $N$ bosonic atoms confined in a one-dimensional chain with $L$ sites. As~depicted in Figure~\ref{potential}a, the~Rydberg-dressed soft-core potential induces interactions  between atoms in different sites. The~dynamics of this model are described by an extended Bose--Hubbard model~\cite{PhysRevA.85.053615} ($\hbar$ = 1):
\begin{equation}
\hat{H} = \sum_{j=1}^{L}\Gamma_j\hat{n}_{j}-J \sum_{\langle i,j\rangle}^{L} (\hat{a}_{i}^{\dagger}\hat{a}_{j}+h.c)
 +\frac{g}{2}\sum_{j=1}^{L}\hat{n}_{j}(\hat{n}_j - 1)+\sum_{i\leq j}^{L}\Lambda_{i,j}\hat{n}_{i}\hat{n}_{j},
\label{Ham}
\end{equation}
where $\hat{a}_{j}$ and $\hat{a}_{j}^{\dagger}$ are bosonic annihilation and creation operators acting on site $j$, with~the corresponding number operator $\hat{n}_j=\hat{a}_{j}^{\dagger}\hat{a}_{j}$.\added{\ Here, $\langle i,j\rangle$ denotes the pair of nearest-neighbour indices.} Parameter $\Gamma_{j}=-[j-1 - \lfloor{L/2}\rfloor]\gamma$  describes a local tilt of the potential well, in~which $\gamma$ represents the level bias between neighboring sites. \emph{J} denotes the hopping rate of atoms between neighboring sites. The~s-wave interaction is characterized by $g=4\pi a_s/m$, with $a_s$ and $m$ being the s-wave scattering length and mass of the atom~\cite{Pethick2008a}. The~long-range soft-core interaction induced by the Rydberg dressing is given by $\Lambda_{i,j}=C_{6}/[|i-j|^6d^6+R_c^6]$, where $C_6$, $d$, and~$R_c$ are an effective dispersion coefficient, lattice constant and soft-core radius, respectively.


 Hilbert space of the EBHM is given by $\mathcal{L}=(N+L-1)!/N!(L-1)!$. When $N\sim L$, it increases rapidly when $N$ is large. For~example,  $\mathcal{L}=92378$ when $N=L=10$.  Resulting numerical calculations become difficult in unit-filling situations. In~this work, we focused on large filling regime \replaced{($N\gg L$)}{($N\ll L$)} in a minimal trapping setting, i.e.,~Rydberg-dressed bosons trapped in triple-well potential ($L=3$). Here, the dimension of the Hilbert space was $\mathcal{L}=(N+1)(N+2)/2$, which is quadratic with $N$. This allowed us to deal with large particle numbers $N\sim 100$. Due to the high filling $N/L\gg 1$, we could also directly compare the quantum many-body calculations with mean-field~results. 
 
With the above consideration, we obtained the Hamiltonian of the three-well system,
\begin{equation}
    \begin{split}
        &\hat{H} = \gamma(\hat{n}_{1}-\hat{n}_{3})-J(\hat{a}^{\dagger}_{1}\hat{a}_{2}+\hat{a}_{2}^{\dagger}\hat{a}_{1})-J(\hat{a}^{\dagger}_{2}\hat{a}_{3}+\hat{a}_{3}^{\dagger}\hat{a}_{2})\\
        &+\frac{W}{2N}(\hat{n}_1^{2}+\hat{n}_2^{2}+\hat{n}_3^2)+\frac{U}{N}(\hat{n}_{1}\hat{n}_{2}+\hat{n}_{2}\hat{n}_{3})
        +\frac{V}{N}\hat{n}_{1}\hat{n}_{3},
    \end{split}
    \label{Ham2}
\end{equation}
 where we defined $W=\left(Ng+2N\Lambda_{j,j}\right)$, $U=N\Lambda_{j,j\pm 1}$ and $V=N\Lambda_{j,j\pm 2}$. We neglected constant terms that will not affect many-body dynamics. In~following numerical analysis, we scaled the Hamiltonian with respect to $J$.  For~concreteness, we chose parameters such that the effective on-site interaction vanishes but nearest-neighbor (NN) interaction is strong, i.e.,~$W=0$ and~$U=2V$~\cite{McCormack2020b}. The~latter can be achieved by tuning the s-wave scattering length through Feshbach resonance~\cite{Pethick2008a}. This allows us to concentrate on effects due to the long-range~interactions. 

\section{Results}
\label{result}

Before discussing many-body calculations of the EBHM, we show the signature of chaos of this model in the semiclassical limit~\cite{McCormack2020b,mccormack_hyperchaos_2021}. \added{In the semiclassical (mean-field) calculation, one replaces the bosonic operators $a_i$($a_i^{\dagger}$) with complex classical fields $\psi_i$($\psi_i^{*}$). The~mean-field calculation is relatively simple,  as~the number of complex fields is equal to the number of site $L$. } More details of the semiclassical calculations can be found in Refs.~\cite{McCormack2020b,mccormack_hyperchaos_2021}. A~key figure of merit in the mean-field calculation is Lyapunov exponents. Positive and large Lyapunov exponents indicate the emergence of classical chaos. This means that perturbations to initial states will grow exponentially with time. Examples of Lyapunov exponents are shown in Figure~\ref{potential}b,c. In~Figure~\ref{potential}b, Lyapunov exponent $\lambda$ is close to zero when $U/J<3.5$ (fixing $\gamma/J= 2.5$), where dynamics are linear. Positive $\lambda$ is found at around $U/J \ge 3.5$, and~it reaches its maximum at around $U/J=7.0$. Focusing on this strong interaction region ($U/J=7.0$), we calculated Lyapunov exponents by varying the tilting $\gamma$. We found that $\lambda$ increases with an increasing $\gamma$, and~achieves its maximum at around $\gamma/J=2.5$. 

\begin{figure}[H]

    \includegraphics[scale=0.12]{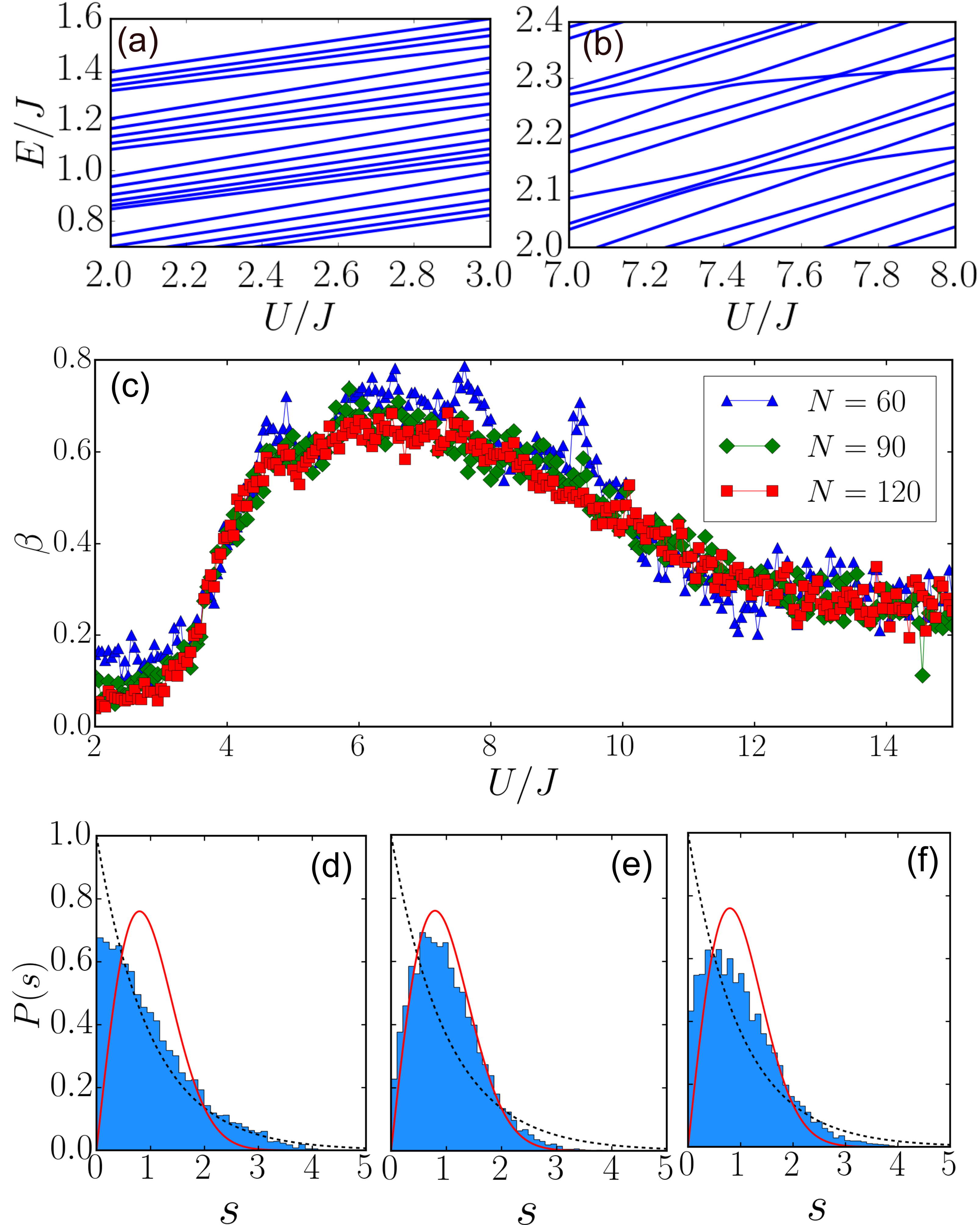}
	\caption{(\textbf{a},\textbf{b}) {Eigenspectrum} as a function of $U/J$. When $U$ is large, avoided crosses are found. In~the calculation, $\gamma/J=2.5$ and $N=12$. (\textbf{c}) Chaos indicator $\beta$ as a function of $U/J$. Maximal $\beta$ is found at around $U/J=7$, where the level spacing approaches the WD distribution. (\textbf{d}--\textbf{f}) show examples of level spacing distributions. In~(\textbf{d}--\textbf{f}), $U/J$ is fixed to be 3, 7 and 13, respectively. The~dashed blue and solid red lines in (\textbf{d}--\textbf{f}) are the Poissonian and WD distribution, respectively. When $U/J=7$, the~level is close to the WD distribution. Other parameters are $\gamma/J=2.5$ and $N=120$ in~(\textbf{d}--\textbf{f}).} 
	\label{beta}
\end{figure}
\begin{figure}[H]

	\includegraphics[scale=0.12]{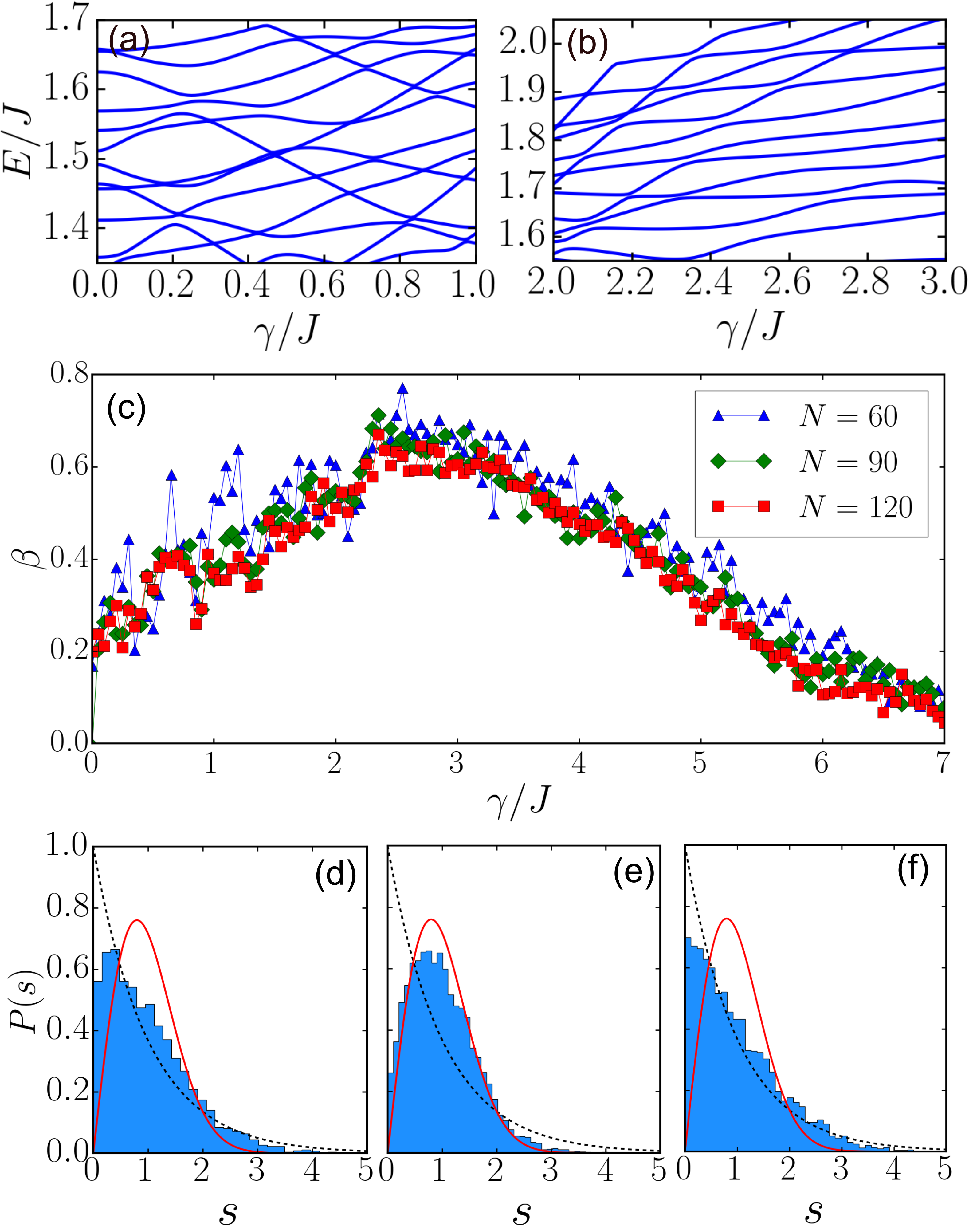}
	\caption{(\textbf{a},\textbf{b}) {Eigenspectra} at different intervals of $\gamma/J$ with $U/J=7$ and $N=12$. Direct and avoided level crossings are found in (\textbf{a}). In~(\textbf{b}), only avoided level crossings are found. (\textbf{c}) Chaos indicator $\beta$ as a function of $\gamma/J$. Around $\gamma/J=2.5$, $\beta$ reaches the maximal value, where the level spacing approaches the WD distribution. (\textbf{d}--\textbf{f}) show examples of level spacing distributions. In~(\mbox{\textbf{d}--\textbf{f})}, $\gamma/J$ is fixed to be 0, 2.5 and 7, respectively. The~dashed blue and solid red lines in \mbox{(\textbf{d}--\textbf{f})}~represent the Poissonian and WD distribution, respectively. When $\gamma/J=2.5$, the~levels approach the WD distribution. Other parameters are $U/J=7$ and $N=120$ in (\textbf{d}--\textbf{f}).}
	\label{beta2}
\end{figure}

We now turn to analyzing static and dynamics properties of the three-well system with finite $N$. By~numerically diagonalizing Hamiltonian (\ref{Ham2}) eigenstates, $|j\rangle\equiv\sum_{\textbf{n}}C_{\textbf{n}}^{j}|\textbf{n}\rangle$ and eigenvalue $E_{j}$ were obtained. Here, $C_{\textbf{n}}^{j}$ is the probability amplitude of Fock state $|\textbf{n}\rangle\equiv|n_{1},n_{2},n_{3}\rangle$ in which $n_i$ denotes the number of particles at the $i$-th site. Statistical distributions of $E_j$ provide valuable information on whether the system is integrable or chaotic~\cite{haake1991quantum}. If~the system is integrable, corresponding energy levels are not correlated, and~are not prohibited from direct crossing when varying parameters~\cite{Berry1977LevelCI,PhysRevLett.74.518, PhysRevA.43.4237},  whereas, in the quantum chaotic region, energy levels are correlated and crossings are avoided~\cite{haake1991quantum,guhr1998random}. In~the weak interaction limit $U\to 0.0$ (i.e., Figure~\ref{beta}a), atom hopping dominates  dynamics such that the system is integrable~\cite{castro2021quantumclassical}. As~we increase $U/J$ to 7.0,  we can see that anti-crossings start to appear, which indicate the emergence of quantum chaos (Figure~\ref{beta}b). When fixing $U/J=7.0$, we show eigen-energies in the interval $0.0<\gamma <1.0$ in Figure~\ref{beta2}a and $2.0<\gamma<3.0$ in Figure~\ref{beta2}b. The~difference is that both direct and avoided crossings are found in Figure~\ref{beta2}a, whereas only avoided crossings are encountered when $\gamma$ is large, as~shown in Figure~\ref{beta2}b.

\unskip

\unskip
\subsection{Level~Statistics}
\label{levelstat}
To understand the different behaviors of the eigen-energies, we analyzed statistics of the nearest-neighbor spacing $s$ between energy levels. The~probability distribution $P(s)$ of the spacing $s$ can be used to distinguish integrable and chaotic dynamics~\cite{haake1991quantum}. For~integrable systems, $P(s)$ follows the Poissonian distribution,
\begin{equation}
    P_{P}(s)=e^{-s}.
\end{equation}

For chaotic systems, their level spacing corresponds to the Gaussian orthogonal ensemble (GOE)~\cite{PhysRevLett.52.1}. Distribution $P(s)$ changes to the Wigner--Dyson (WD) distribution~\cite{10.2307/1970079},
\begin{equation}
P_{WD}(s) = \frac{\pi s}{2}e^{-\pi s^2/4}.
\end{equation}

In realistic situations, the~level spacing will typically not follow the Poissonian or WD distribution exactly. $P(s)$ can be fitted by the Brody distribution~\cite{RevModPhys.53.385},
\begin{equation}
\label{brody}
P_\beta (s) = (\beta+1) b s^{\beta} \exp(-b s^{\beta+1}), 
\end{equation}
where $\beta$ is a chaos indicator and $b$ depends on $\beta$,
\begin{equation}
	b =\left[ \Gamma \left( \frac{\beta+2}{\beta+1} \right) \right]^{\beta+1},
\end{equation}
where $\Gamma(x)$ is the gamma function. Here, $0 \leq \beta \leq 1$ measures the repulsion of the levels, also known as the level repulsion exponent. For~chaotic systems, $\beta \sim 1$, where the Brody distribution recovers the WD profile, whereas  $\beta \sim 0$ in integrable systems, where the Brody distribution becomes the Poissonian~distribution.

After carrying out the standard unfolding procedure~\cite{haake1991quantum},  the~level statistics were calculated and fitted with Equation~(\ref{brody}). Values of $\beta$ were extracted and are shown in Figure~\ref{beta}c and Figure~\ref{beta2}c, respectively. Fixing $\gamma/J=2.5$, it was found that $\beta$ is small when $U/J\to 0.0$. In~this region, the level statistics follow the Poissonian distribution, as~illustrated in Figure~\ref{beta}d. Then, $\beta$ keeps increasing with am increasing $U/J$ until \mbox{$U/J\sim 7.0$}, where the level statistics display the WD distribution i.e.,~Figure~\ref{beta}e. Hence, the system enters the quantum chaos regime. Further increasing $U$, $\beta$ decreases gradually, where the statistics follow the Poissonian distribution again, i.e.,~Figure~\ref{beta}f. The~results show that our model is largely integrable in the weakly and strongly interacting regimes, which is similar to the Bose--Hubbard model~\cite{Kolovsky_2004}.  The~trend of $\beta$ shown in Figure~\ref{beta}c is consistent with the Lyapunov exponent shown in~Figure~\ref{potential}b.

 In Figure~\ref{beta2}c, we plot $\beta$ as a function of $\gamma/J$. Similar to the Lyapunov exponents shown in Figure~\ref{potential}c, the~maximal value of $\beta$ is found around $\gamma/J\approx 2.5$, where dynamics are strongly quantum chaotic. Values of $\beta$ decrease quickly when $\gamma/J\to 0$ or  $\gamma/J\gg 2.5$, where the triple-well system becomes integrable. Examples of corresponding level distributions are shown in Figure~\ref{beta2}d--f. The~levels approximately follow a Poissonian distribution when $\gamma/J\to 0$, i.e.,~Figure~\ref{beta2}d, or $\gamma/J\gg2.5$, i.e.,~Figure~\ref{beta2}f, whereas they follow a WD distribution when $\gamma/J =2.5$, i.e.,~Figure~\ref{beta2}e.

\subsection{Entanglement~Entropy}
\label{eentropy}
In this section, we analyze the entanglement entropy of our model. The~three wells were divided into subsystems of the left well (part $A$) and the other two wells (part $B$). Entanglement entropy with respect to subsystem $A$ is given by~\cite{vonNeumann1996}
\begin{equation}
	S_{EE} = -\text{Tr}(\rho_{A}\ln{\rho_{A}}),
	\label{vnentropy}
\end{equation}
where $\rho_A=\text{Tr}_{B}(\rho)$ is the reduced density matrix by tracing out the subsystem $B$ from the density matrix $\rho=|\psi \rangle\langle\psi|$ of the whole system.  Entanglement entropy $S_{EE}$ measures how much subsystem $A$ and $B$ are intertwined with each other. For~the three-well bosonic system, it has a maximal value $S_{EE}^{\text{max}}=\ln (N+1)$, which only depends on the total number of atoms of the whole system (see Appendix \ref{app} for calculations). 

To evaluate the entanglement entropy numerically, we started with an initial {state} 
 $ |\psi_{0}\rangle = \sum_{\textbf{n}}a_{\textbf{n}}|\textbf{n}\rangle$. The~time-dependent many-body state $|\psi(t)\rangle=e^{-i\hat{H}t}|\psi_{0}\rangle$ was expanded in the Fock~basis,
\begin{equation}
	|\psi(t)\rangle=\sum_{j}\sum_{\textbf{n}}e^{-iE_{j}t}\alpha_{j}C_{\textbf{n}}^{j}|\textbf{n}\rangle,
	\label{evolestateexpanded}
\end{equation}
in which we defined $\alpha_{j}\equiv \sum_{\textbf{n}}(C_{\textbf{n}}^{j})^{*}a_{\textbf{n}}$. Given the time-dependent many-body state in the Fock basis, the~entanglement entropy $S_{EE}(t)$ can be~evaluated.

\begin{figure}[H]

	\includegraphics[width=0.65\linewidth]{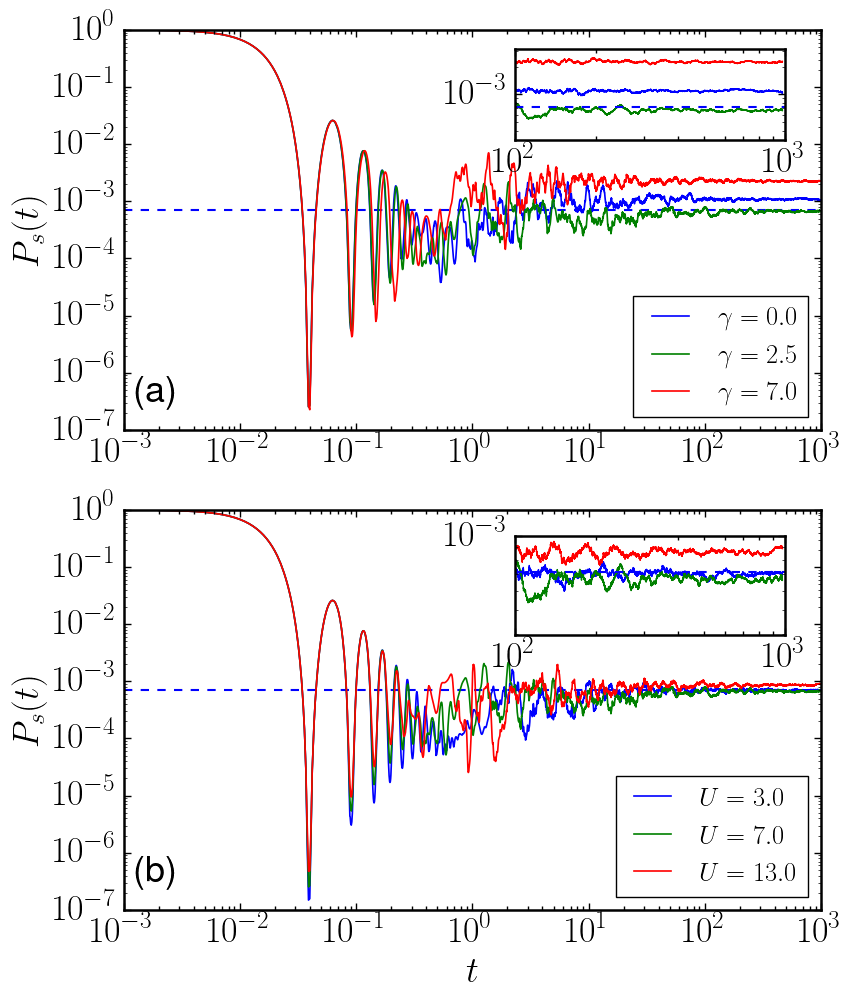}
	\caption{Moving averages of survival probabilities. We set the temporal window size to be $0.04/J$ in numerical calculations. $U/J$ was fixed to be 7.0 in (\textbf{a}). $\gamma/J$ was fixed be 2.5 in (\textbf{b}). \replaced{The dashed blue lines in both panels denote the saturation value of survival probability of GOE matrices.\ }{The dashed blue lines in both panels denote the survival probability of GOE matrices.} The insets of both panels show the moving averages of survival probabilities beyond $\text{log}_{10}(t) = 2$. In~both panels, $N=90$.}
	\label{figsps}
\end{figure}

In Figure~\ref{figsps}a, the dynamical evolution of $S_{EE}(t)$ is shown for $U/J = 3.0$, $7.0$ and $13.0$, with initial state $|N/3,N/3,N/3\rangle$ and $N=120$. The~entropy initially increases monotonically for a short period of time and~then fluctuates around a constant value with small amplitudes. Here, the saturation values of $S_{EE}(t)$ depend on $U/J$. If~one increases $U$, amplitudes of the fluctuation will be further reduced. To~explore the dependence of the entropy on parameters, we numerically calculated time-averaged and normalized entanglement entropy, $\bar{S}_{EE}=1/(S_{EE}^{\text{max}}\Delta t)\int_{t_i}^{t_f} S_{EE}dt$, where $\Delta t=t_f-t_i$. The~normalization could partially eliminate the dependence of $\bar{S}_{EE}$ on $N$. In~the numerical calculation, we chose $t_i=4/J$ to avoid the influence due to the initial stage of the entropy.  We set $t_f=20/J$ in calculations, as~$S_{EE}$ saturates when $t_f\le 20/J$.
 \added{It is important to note that long-time behaviors of $S_{EE}$ are independent of the initial configurations, though~details of the dynamics depend on the configuration. By~focusing on $|N/3, N/3, N/3\rangle$ initially, we can compare dynamics of various parameters. This is particularly useful when $\gamma=0$.}
\vspace{-6pt}
\begin{figure}[H]

	\includegraphics[width=8.5cm, height=8.5cm]{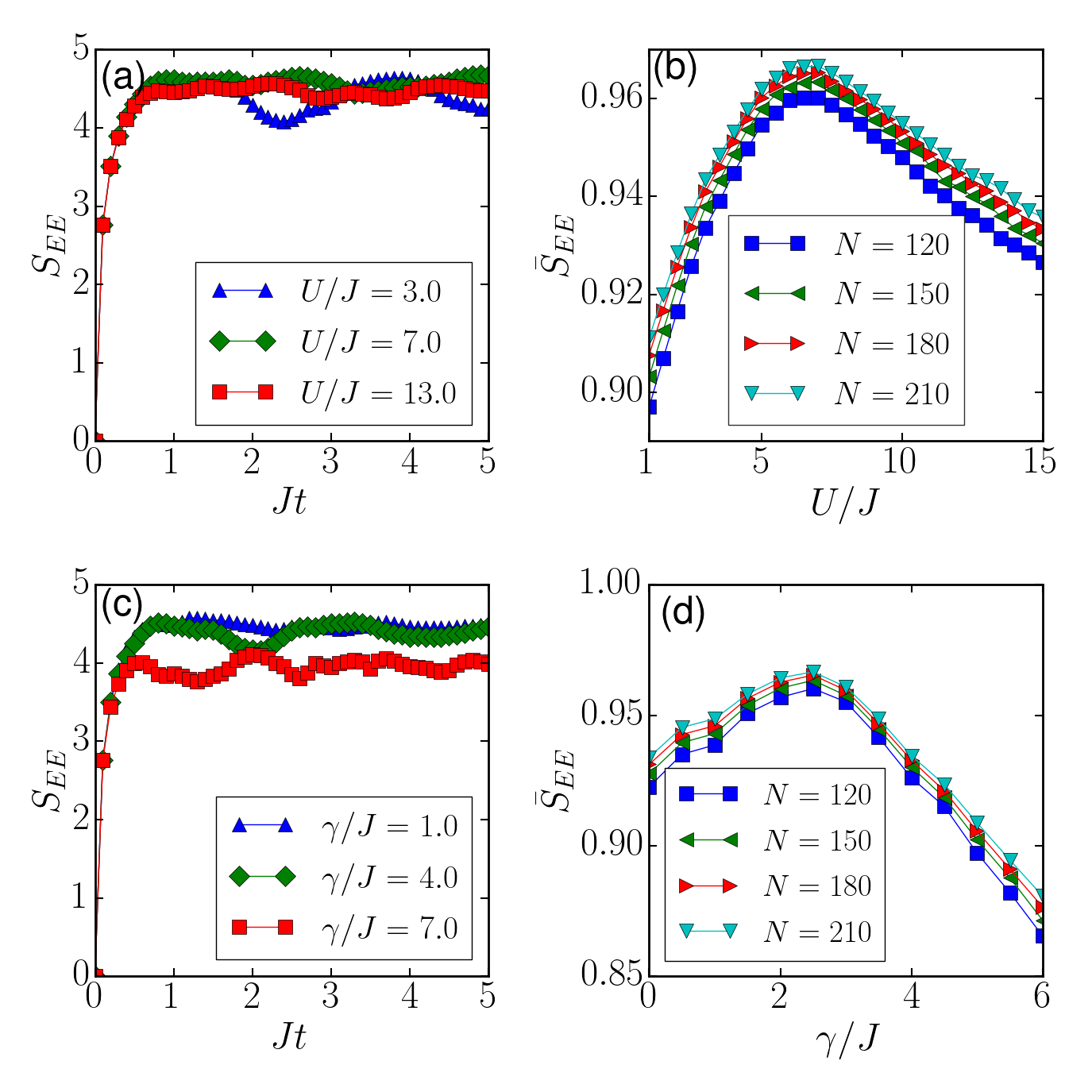}
	\caption{(\textbf{a}) Dynamical evolution of entanglement entropy for $U/J=3.0$, $7.0$ and $13.0$. (\textbf{b}) Normalized average entropy $\bar{S}_{EE}$ as a function of $U/J$ for different $N$.  (\textbf{c}) Dynamical evolution of $S_{EE}$ for  $\gamma/J=1.0$, $4.0$ and $7.0$. (\textbf{d}) $\bar{S}_{EE}$ with respect to $\gamma/J$ for different $N$. In~(\textbf{a},\textbf{b}) panels, $\gamma/J = 2.5$. In~ (\textbf{c},\textbf{d}) panels, $U/J$ is fixed to be~7.0.}
	\label{evolEEs}
\end{figure}

The results of $\bar{S}_{EE}$ as a function of $U$ are shown in Figure~\ref{figsps}b, where $\gamma/J$ was fixed to be 2.5. It is small when $U/J\ll 7.0$. $\bar{S}_{EE}$ then increases with $U$ and gains its peak value at $U/J\approx 7.0$. The~peak value is close to the upper bound $S_{EE}^{\text{max}}$ when $N$ is large (e.g., $N>120$). Further increasing $U$, $\bar{S}_{EE}$ decreases gradually. This leads to a similar pattern to $\beta$, i.e.,~Figure~\ref{beta}c. Importantly, we note that the maximal $\bar{S}_{EE}$ appears at the same $U/J$ as that of $\beta$.  Hence, both predict the position of quantum chaos as indicated by level statistics, i.e.,~Figure~\ref{beta}. We note that $\bar{S}_{EE}$ is shifted upwards (increased) globally when $N$ is increased. Such a dependence is different from $\beta$, where increasing $N$ will suppress the fluctuation and~lead to a better~fitting. 

In Figure~\ref{figsps}c,d, we show the dependence of the entanglement entropy on $\gamma$ while fixing $U/J=7.0$. The~entanglement entropy first increases rapidly with time and then saturates when $Jt\gtrsim 4.0$, as~depicted in Figure~\ref{figsps}c. The~saturation varies with $\gamma$. To~explore this dependence, we evaluated the normalized average entropy $\bar{S}_{EE}$, and~show the results in Figure~\ref{figsps}d. $\bar{S}_{EE}$ first increases with $\gamma$, reaches peak values around $\gamma/J=2.5$ and~then decreases with an increasing $\gamma$. Such a pattern is similar to the dependence of $\beta$ on $\gamma$, as~seen in Figure~\ref{beta2}c. Increasing atom number $N$, the~profile of $\bar{S}_{EE}$ is largely unchanged, but~is shifted globally~upwards.

\subsection{Survival Probability and Variance of~Populations}
\label{surProb}
The survival probability $P_s(t)=|\langle \psi_{0} | \psi(t) \rangle|^2$ gives the probability of finding  initial state $|\psi_{0}\rangle$ at time $t$.  
In the long time limit,  $P_s(t)$ will stay in close proximity to that of GOE matrices if~the dynamics are chaotic. \replaced{The survival probability of GOEs saturates at $P_s^{\text{GOE}} \approx 3/D$~\cite{PhysRevB.99.174313} at the long time limit,
}{The minimal value of survival probability from GOE matrices is $P_s^{\text{GOE}} \approx 2/D$~\cite{PhysRevB.99.174313},} which depends only on the dimension $D$ of the Hilbert space. This provides a way to characterize chaotic quantum dynamics. Considering initial state $|\textbf{n}_0 \rangle=|N/3,N/3,N/3\rangle$ ($N=90$), the~respective survival probability is
\begin{equation}
    \label{modelsp}
    P_s(t) = \left|\sum_{j}e^{-iE_jt}\left|C_{\mathbf{n}_0}^j\right|^2\right|^2,
\end{equation}
where $C_{\mathbf{n}_0}^j$ denotes the probability amplitude of the initial state $|\mathbf{n}_{0}\rangle$.

In Figure~\ref{evolEEs}, we plot the moving average of the survival probability within temporal windows of constant size together with \replaced{the saturation value}{\ the minimal value} of the survival probability of GOE matrices with $D=4186$. From~Figure~\ref{evolEEs}a, it can be seen that the long time limit  of the survival probability at $\gamma/J=2.5$ is in close proximity to the GOE.  We can also see that the system significantly deviates from GOE when  $\gamma/J$ is away from $2.5$.  The~evolution of the survival probability for different $U$ is shown in Figure~\ref{evolEEs}b. In~the long time limit, the~survival probability approaches that of the GOE when $U\approx 7.0$. \added{This seemingly contradicts the prediction based on $\beta$ and entropy. However, we note that the system is already in the chaotic regime with the two parameters. It is not so surprising that the long-time survival probability is so close. On~the other hand, the~survival probabilities are generally close to each other in all the cases (note the logarithmic scale in {Figure}~\ref{evolEEs}), where values of the survival probability are very small. These observation means that they are difficult to observe in cold atom experiments. In~the following, we will show that variances of populations in different potential wells can be used to identify chaotic dynamics. We will show that the population variance exhibits different behaviors in the chaotic region by varying parameter $U$ and $\gamma$. Changes in the variance are sizable, which might provide a plausible way to probe signatures of quantum chaos in the triple-well system. }\vspace{-6pt}

\label{pop}
\deleted{The survival probabilities are generally close to each other in all different cases. In~addition, values of the survival probability are very small. These observation means that they are difficult to observe in cold atom experiments. In~the following, we will show that variances of populations in different potential wells can be used to identify chaotic dynamics. We will show that the population variance exhibits different behaviors in the chaotic region by varying parameter $U$ and $\gamma$. Changes in the variance are sizable, which might provide a plausible way to probe signatures of quantum chaos in the triple well system. }

\begin{figure}[H]

	\includegraphics[scale=0.070]{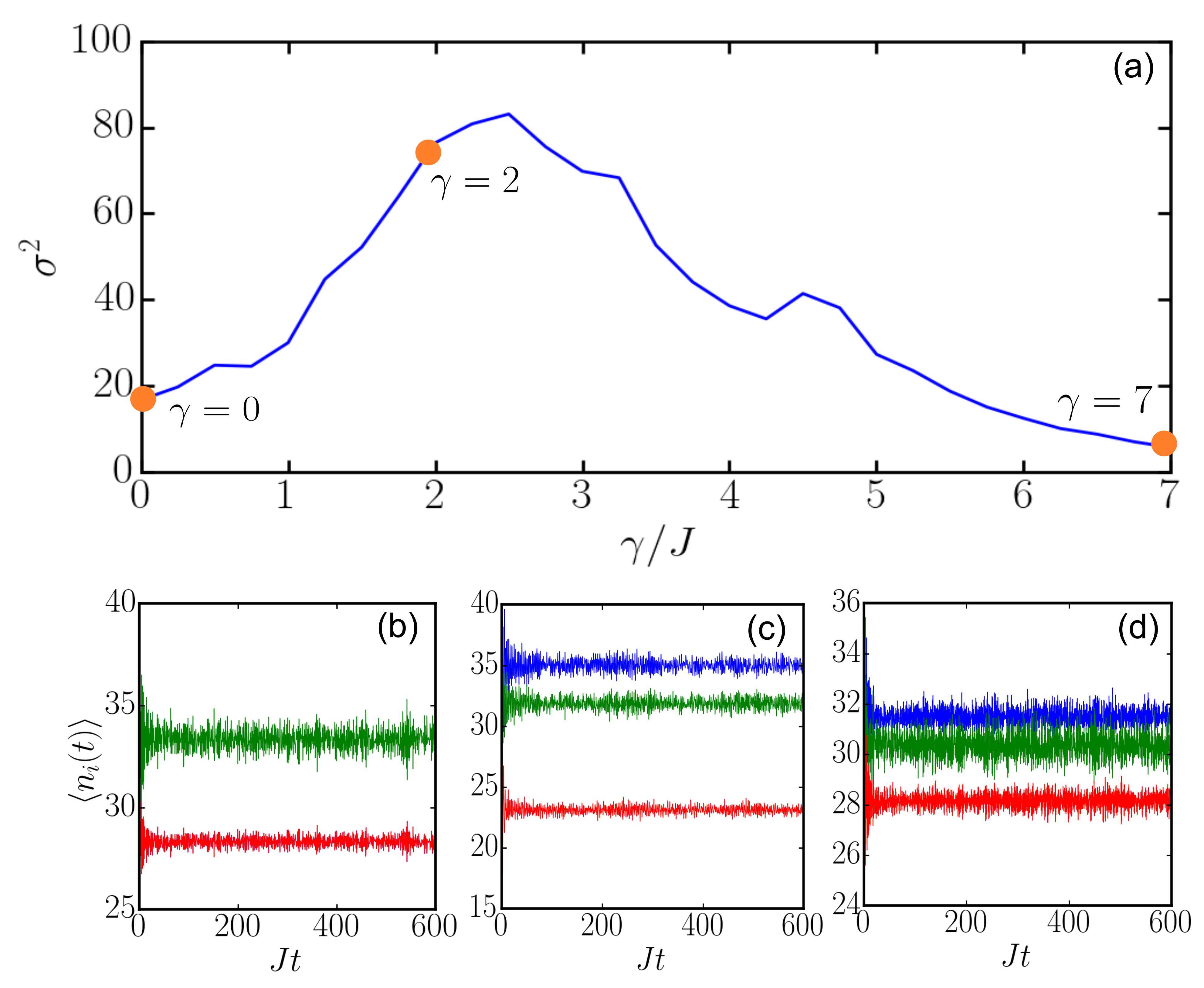}
	\caption{(\textbf{a}) Population variances with respect to $\gamma$. Evolution of population when $\gamma/J =0$ (\textbf{b}), $\gamma/J = 2$ (\textbf{c}) and $\gamma/J = 7$ (\textbf{d}). In~these panels, blue, green and red lines denote the expectation values in the leftmost,  middle and rightmost site, respectively. In~all panels, we choose $U/J=7.0$ and~$N=90$.}
	\label{siteExpectGamma}
\end{figure}

\begin{figure}[H]

	\includegraphics[scale=0.070]{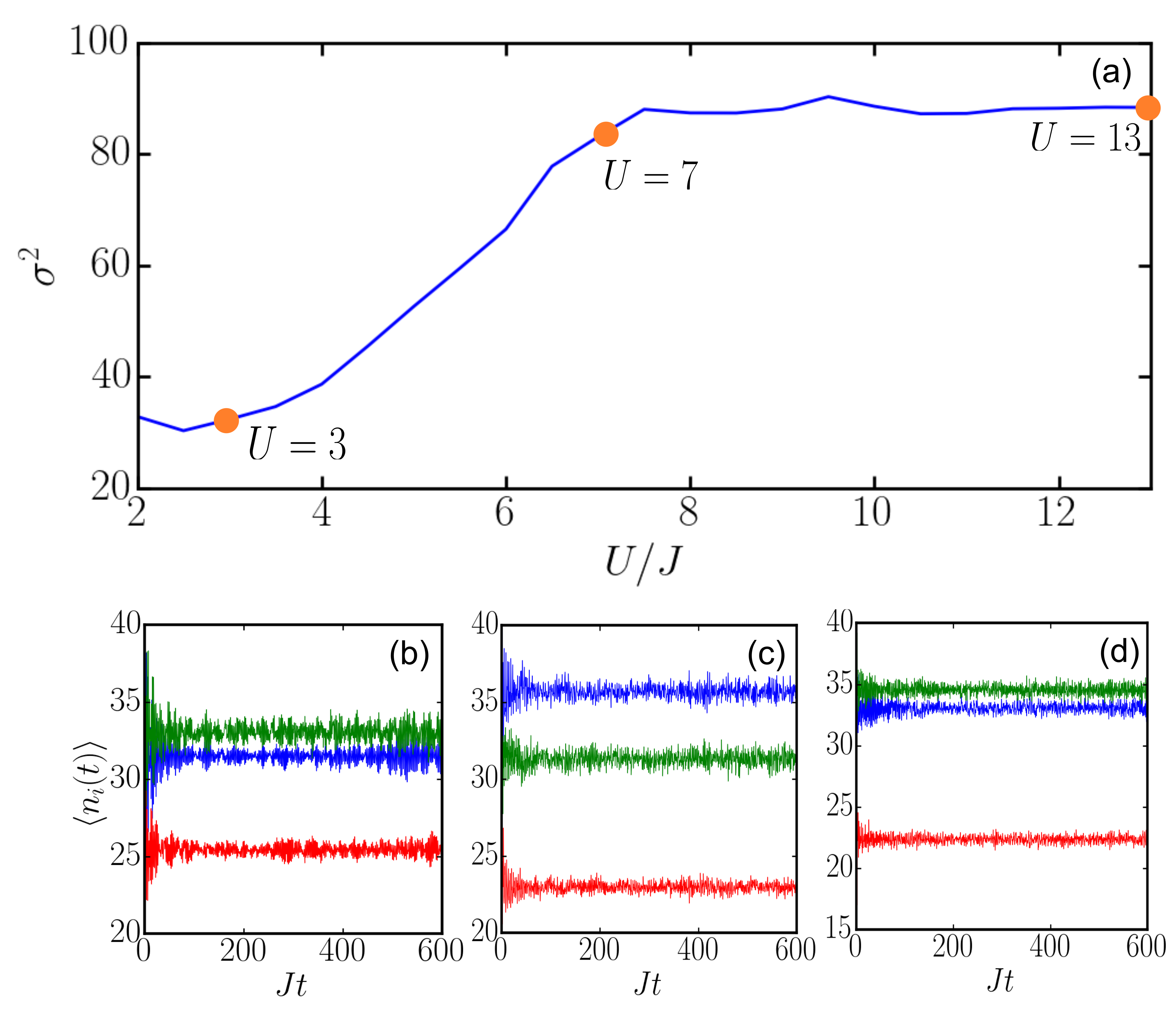}
	\caption{(\textbf{a}) Population variances with respect to $U$. Evolution of populations when $U=3$ (\textbf{b}), $U=7$ (\textbf{c}) and $U=13$ (\textbf{d}). In~each panel, blue, green and red lines denote the population  in the leftmost, middle and rightmost site, respectively. In~the calculation, $\gamma=2.5$ and $N=90$.}
	\label{siteExpectU}
\end{figure}

\textls[-30]{We numerically evaluated population $\langle n_i\rangle$ in the $i$-th site with initial state  $|N/3, N/3, N/3 \rangle$.} The time-averaged variance $\sigma^2$ of the populations in the three wells is defined as
\begin{equation}
    \sigma^2 = \sum_{i}^{3} \left(\frac{1}{\Delta t}\int_{t_{0}}^{t_{\text{f}}}  \langle n_i(t)\rangle dt - \frac{N}{3} \right)^2,
    \label{variance}
\end{equation}
where $\Delta t = t_{\text{f}}-t_0$. Figure~\ref{siteExpectGamma} shows dependences of the variance on $\gamma$ and with $U/J=7.0$. Here, $\sigma^2$ first increases, gains a peak value at around $\gamma\approx 2.5$ and~then decreases with an increasing $\gamma$. This dependence is similar to that of the Lyapunov exponent, i.e.,~Figure~\ref{potential}c, chaos indicator, i.e.,~Figure~\ref{beta2}c, and entanglement entropy, i.e.,~Figure~\ref{evolEEs}d. Hence, the similar profiles indicate that one could measure the variance to identify chaotic dynamics. Furthermore, we show the variance as a function of $U/J$ in Figure~\ref{siteExpectU}. We find that the variance first increases with an increasing $U/J$ and~then saturates at around $90$. Hence, the variance becomes insensitive if we further increase $U$. This could be attributed to the fact that strong interactions suppress hopping and~hence reduce the population variance. Though~the profile is different from other chaos indicators in this case, we notice that the saturation starts from $U/J=7$. This corresponds to the parameter where other chaos indicators arrive at their maximal~values. 

\unskip
\unskip
\section{Discussion and~Conclusions}

\label{conclusion}
In this work, we studied quantum dynamics of Rydberg-dressed bosons within a triple-well setup. In~the semiclassical regime, we identified parameter regions where Lyapunov exponents are large. Through diagonalizing the Hamiltonian, we showed that the level statistics fulfil the Wigner--Dyson distribution in the parameter region where the Lyapunov exponents are maximal, which indicates the emergence of quantum chaos in this system. By~analyzing the quantum many-body dynamics, we showed that time-averaged entanglement entropy has a similar dependence on $U$ and $\gamma$.  We showed that the survival probability exhibits a similar dependence on $U$ and $\gamma$ to that of the level statistics and Lyapunov exponents. Our numerical calculations show that the time-averaged variance of the population could be used to provide chaotic~dynamics.

\vspace{6pt} 



\authorcontributions{Numerical calculation, T.Y. and M.C.; formal analysis, T.Y. and W.L.; {writing---original} draft preparation, T.Y. and M.C.; writing---review and editing, R.N. and W.L.; supervision, W.L. All authors have read and agreed to the published version of the~manuscript.}

\funding{This research is partially funded by the EPSRC through Grant Nos.~EP/W015641/1 and ~EP/W524402/1.}

\dataavailability{The data generated for this paper are available with doi: \url{https://doi.org/10.5281/zenodo.7615623}.}

\acknowledgments{T.Y. would like to thank T. Hamlyn for useful discussions and critical reading of the draft.\added{\ We thank Lea Santos for a critical comment on the saturation value in Figure~\ref{evolEEs}.}}

\conflictsofinterest{The authors declare no conflict of~interest.}  





\appendixtitles{yes} 
\appendixstart
\appendix

\section{Saturation Values of \boldmath{$S_{EE}$}}\label{app}

In this appendix, we derive the maximal entanglement entropy following Ref.~\cite{PhysRevB.101.060401}. Given a bosonic system bipartited into two subsystems A and B, the~full Hilbert space $\mathcal{H}$ can be written in the form
\begin{equation}
	\mathcal{H}=\bigoplus_{n_{A}=0}^{N}\mathcal{H}_{A}^{(n_A)}\otimes \mathcal{H}_{B}^{(N-n_A)},
\end{equation}
in which $n_A$ and $n_B$ are numbers of particles in subsystem $A$ and $B$.

A quantum state $|\psi\rangle$ can be expanded using the Fock basis,
\begin{equation}
	\label{generalpsi}
	|\psi\rangle = \sum_{n_A=0}^{N}\sum_{i^{(n_A)}=1}^{d_{A}^{(n_A)}}\sum_{j^{(N-n_A)}=1}^{d_{B}^{(N-n_A)}}a_{ij}^{(n_A)}|i^{(n_A)}\rangle\otimes |j^{(N-n_A)}\rangle,
\end{equation}
in which $d_{A}^{(n_A)}$ and $d_{B}^{(N-n_A)}$ denote the dimensions of $\mathcal{H}_{A}^{(n_A)}$ and $\mathcal{H}_{B}^{(N-n_A)}$, respectively, and $\{|i^{(n_A)}\rangle\}_{i=1}^{d_A^{(n_A)}}$ and $\{|j^{(N-n_A)}\rangle\}_{j=1}^{d_B^{(N-n_A)}}$ form the computational basis of $\mathcal{H}_{A}^{(n_A)}$ and $\mathcal{H}_B^{(N-n_A)}$, respectively.

For the sake of simplicity, we can project $|\psi\rangle$ into a certain sector of $n_A$ forming $|\psi_{n_A}\rangle$ so that we do not need to consider the first summation in Equation~(\ref{generalpsi}) and we have $|\psi_{n_A}\rangle$ in the form
\begin{equation}
	|\psi_{n_A}\rangle = \sum_{i^{(n_A)}=1}^{d_{A}^{(n_A)}}\sum_{j^{(N-n_A)}=1}^{d_{B}^{(N-n_A)}}a_{ij}^{(n_A)}|i^{(n_A)}\rangle\otimes |j^{(N-n_A)}\rangle.
\end{equation}

By applying the singular value decomposition (SVD) onto the $d_{A}^{(n_A)}\times d_{B}^{(N-n_A)}$ matrix $[a^{(n_A)}]$, we can rewrite such a matrix into the form
\begin{equation}
	\label{specificpsi}
	[a^{(n_A)}] = U\Sigma V^{\dagger},
\end{equation}
in which $U$ is a $d_{A}^{(n_A)}\times d_{A}^{(n_A)}$ semi-unitary matrix, $\Sigma$ is a $d_{A}^{(n_A)}\times d_{B}^{(N-n_A)}$ diagonal matrix with non-negative real numbers and $V$ is a $d_{B}^{(N-n_A)}\times d_{B}^{(N-n_A)}$ semi-unitary matrix. After~applying Schmidt decomposition, Equation~(\ref{specificpsi}) can be rewritten as
\begin{equation}
	|\psi_{n_A}\rangle = \sum_{i=1}^{d^{(n_A)}} \alpha_{i}^{(n_A)}|u_i^{(n_A)}\rangle\otimes |v_{i}^{(N-n_A)}\rangle,
\end{equation}
in which $d^{(n_A)}=\text{min}\{d_A^{(n_A)},d_B^{(N-n_A)}\}$, $\{\alpha_{i}^{(n_A)}\}_{i=1}^{d^{(n_A)}}$ are elements of $\Sigma$, and~$\{|u_{i}^{(n_A)}\rangle\}_{i=1}^{d^{(n_A)}}$ and $\{|v_{i}^{(N-n_A)}\rangle\}_{i=1}^{d^{(n_A)}}$ are the first $d^{(n_A)}$ columns of matrices $U$ and $V$, respectively.

\textls[-35]{The entanglement entropy $S_{EE}^{(n_A)}$ with respect to $n_A$-sector can therefore be easily~calculated:}
\begin{equation}
	\label{trueEE}
	S_{EE}^{(n_A)} = -\sum_{i=1}^{d^{(n_A)}} |\alpha_{i}^{(n_A)}|^2 \text{ln}|\alpha_{i}^{(n_A)}|^2.
\end{equation}

The total entanglement entropy $S_{EE}$ can be computed by summing over all possible~sectors:
\begin{equation}
	\begin{split}
		S_{EE} &= \sum_{n_A=0}^{N}S_{EE}^{(n_A)}\\
		&= -\sum_{n_A=0}^{N}\sum_{i=1}^{d^{(n_A)}} |\alpha_{i}^{(n_A)}|^2 \text{ln}|\alpha_{i}^{(n_A)}|^2.
	\end{split}
	\label{totalentropy}
\end{equation}

By applying Jensen's theorem onto Equation~(\ref{totalentropy}), the~upper bound of $S_{EE}$ can be found~\cite{PhysRevB.101.060401}:
\begin{equation}
	\label{EEupperbound}
	S_{EE} \leq \text{ln}\sum_{n_A=0}^{N}d^{(n_A)}.
\end{equation}

For a bosonic system with $L$ sites in total and $L_A$ sites in subsystem A, Equation~(\ref{EEupperbound}) can be rewritten:
\begin{equation}
	S_{EE} \leq \text{ln}\sum_{n_A=0}^{N}\text{min}\left\{\binom{L_A+n_A-1}{n_A}, \binom{L-L_A+N-n_A-1}{N-n_A}\right\}.
\end{equation}

In our triple-well setup, subsystem A only contains the first site. Therefore the upper bound of entanglement entropy only depends on the total number of particles:
\begin{equation}
	\label{EEuppberboundspecific}
	S_{EE} \leq \text{ln} (N+1)
\end{equation}

\begin{adjustwidth}{-\extralength}{0cm}
\printendnotes[custom] 

\reftitle{References}

\PublishersNote{}
\end{adjustwidth}
\end{document}